\documentclass[aps,pra,preprint]{revtex4}
\usepackage{amsmath}
\usepackage{epsfig,amssymb,amsfonts}
\usepackage{graphicx}

\begin{document}

\title{AHARONOV-BOHM EFFECT AS A DIFFUSION PHENOMENON}

\author{Charalampos Antonakos}
\affiliation{Department of Physics, School of Natural Sciences, University of Patras, 265 04 Patras, Greece}

\author{Andreas F. Terzis}
\email{afterzis@upatras.gr}
\affiliation{Department of Physics, School of Natural Sciences, University of Patras, 265 04 Patras, Greece}

\today

\begin{abstract}
This paper presents a hydrodynamical view of the Aharonov-Bohm effect, using Nelson's formulation of quantum mechanics. Our aim is to compare our results with other systems and gain a better understanding of the mysteries behind this effect, such as why the motion of a particle is affected in a region where there is no magnetic field. Some theories suggest that this effect is due to the non-local action of the magnetic field on the particle, or even the physical significance of vector potentials over magnetic fields. Our main purpose is to use Nelson's formulation to describe the effect and demonstrate that it can be explained by the direct action of the current surrounding the magnetic field region (i.e. a cylinder) on the particle outside of it. In this context, magnetic fields and vector potentials serve as tools for finding other fundamental quantities that arise from the interaction between two fields: the quantum background fields described by Nelson's quantum theory. Finally, we investigate the relationship between hidden variables and quantum fluctuations and their role in this phenomenon.
\end{abstract}
\maketitle

\section{Introduction}

In 1959, Aharonov and Bohm published a historic article \cite{ahaBohm1959} which highlighted the significance of electromagnetic potential in the quantum domain. They suggested possible experiments in the quantum regime where e/m potentials were necessary for accurately describing reported phenomena. This was in contrast to the classical electromagnetic theory where vector and scalar potentials had only been used as convenient mathematical tools for calculating real fields. 

The observed behavior was not unexpected in quantum mechanics, as the Schrödinger equation is derived from the canonical formalism which requires both the fields and potentials to be described accurately. This indicates that in quantum physics, the fundamental physical entities are the potentials, rather than the fields alone.

An attempt to explain this effect  based on the hydrodynamic description of quantum mechanics, suggested in 1927 by Madelung \cite{madelung}, has appeared in the early 80's by Takabayasi \cite{takabayasi1}.

In this work, we aim to provide a comprehensive description of the Aharonov-Bohm effect by employing an approximation that is closer to a stochastic description \cite{nelson1966derivation}.

\section{Nelson's quantum mechanics in the Aharonov-Bohm effect}

In his historic 1966 article, "Derivation of the Schrödinger equation from Newtonian mechanics", Nelson explored the hypothesis that every elementary particle influenced by an external field behaves as if subjected to Brownian motion, with a diffusion coefficient proportional to Planck's constant and inversely proportional to the particle's mass, and without experiencing any friction \cite{nelson1966derivation}. In 2003, Mita also arrived at similar results without assuming stochastic processes \cite{mita2003dispersive}. In this paper, we adopt a methodology that combines aspects of both approximations to develop a theoretical framework for investigating and understanding the Aharonov-Bohm effect.

The first important physical quantity is the integrand of the momentum operator, i.e.  
\begin{equation} \label{eq:1}
 \Psi^*\hat{P}\Psi=M\left(\overrightarrow{J}+i\overrightarrow{D}\right) 
 \end{equation}
where $\Psi$ is the wave function of a particle with mass M. 
In the rhs of eq. (\ref{eq:1}), $\overrightarrow{J}$ is the probability current density and $\overrightarrow{D}$ is the diffusion current.

Hence,

$$\overrightarrow{J}=\frac{1}{2M}\left(\Psi^*\hat{P}\Psi-\Psi\hat{P}\Psi^*\right)$$

As the current velocity is $\overrightarrow{\eta}=\overrightarrow{J} / \rho$, by describing the wave function in the polar form (i.e. $\psi=e^{R+iS}$), we get 

\begin{equation} \label{eq:2}
    \overrightarrow{\eta}=\frac{1}{2M\rho}\left(\Psi^*\hat{P}\Psi-\Psi\hat{P}\Psi^*\right)=\beta^2\overrightarrow{\nabla}(2S)=\frac{\hbar}{M}\overrightarrow{\nabla}S
\end{equation}
where $\beta^2=\frac{\hbar}{2M}$ is the diffusion coefficient.
Similarly, from the diffusion current we find the dispersive velocity ($\overrightarrow{\xi}=\overrightarrow{D} / \rho$), 

\begin{equation} \label{eq:3}
\overrightarrow{\xi}=\frac{1}{2Mi\rho}\left(\Psi^*\hat{P}\Psi+\Psi\hat{P}\Psi^*\right)=-\beta^2\frac{\overrightarrow{\nabla}\rho}{\rho}=-\beta^2\overrightarrow{\nabla}(2R)=-\overrightarrow{\zeta}
\end{equation}

These last two quantities we derived ($\overrightarrow{\eta}$ and $\overrightarrow{\zeta}$) are in correspondence with $\overrightarrow{v}(\overrightarrow{x},t)$ and $\overrightarrow{u}(\overrightarrow{x},t)$ introduced by Nelson and are called current velocity and osmotic velocity (for a Brownian particle) respectively \cite{nelson1966derivation}. 
For those velocity fields we have that

$$\overrightarrow{v}(\overrightarrow{x},t)=\frac{1}{2}(D_+ + D_-)\overrightarrow{x}(t)$$
and
$$\overrightarrow{u}(\overrightarrow{x},t)=\frac{i}{2}(D_+ - D_-)\overrightarrow{x}(t)$$

where $D_+ \overrightarrow{x}(t)$ and $D_- \overrightarrow{x}(t)$ are regarded as mean forward and mean backward derivative according to Nelson. If we compare these equations from classical Brownian motion (the $i$, however, does not exist in the expression of the classical osmotic velocity, but it exists in the expression of the quantum osmotic velocity) with Mita's equations we can say that

$$D_+(\overrightarrow{x},t)=\frac{\Psi^*\hat{P}\Psi}{M\rho}$$
and
$$D_-(\overrightarrow{x},t)=-\frac{\Psi\hat{P}\Psi^*}{M\rho}$$

Nelson derived the Schrödinger equation using these velocities. In our study, we aim to use these velocities to explore the Aharonov-Bohm effect and gain a better understanding of it. Of course, in our description we will be referring to both osmotic velocity fields and dispersive velocities, since their only difference is when it comes to the sign. In fact, in a quantum system, the existence of the dispersive velocity guarantees the existence of the osmotic velocity and vice versa. 

Let us suppose that a charged particle is in an area without a magnetic field but with a vector potential. In this scenario, the momentum changes, and we have $\hat{P}'=\hat{P}-\frac{q}{C}\overrightarrow{A}$. This change in momentum will have an effect on the momentum density, which can be expressed as follows:
\begin{equation} \label{eq:4}
 \Psi^*\hat{P}'\Psi=M\left(\overrightarrow{\Gamma}_\lambda+i\overrightarrow{\Delta}_\lambda\right)
\end{equation}

We name $\overrightarrow{\Gamma}_\lambda$ and $\overrightarrow{\Delta}_\lambda$ the quasi-probability current and quasi-diffusion current, respectively. For these two quantities, we have:
\begin{equation} \label{eq:5}
\overrightarrow{\Gamma}_\lambda=\frac{1}{M}Re\left(\Psi^*\hat{P}'\Psi\right)
\end{equation}
and
\begin{equation} \label{eq:6}
\overrightarrow{\Delta}_{\lambda}=\frac{1}{M}Im\left(\Psi^*\hat{P}'\Psi\right)
\end{equation}

In general, these two variables will have properties of the traditional probability current and diffusion current. Now, when it comes to the dispersive velocity we have:

\begin{equation} \label{eq:7}
\overrightarrow{\xi}_\lambda=\frac{1}{2Mi\rho}\left(\Psi^*\left(\hat{P}-\frac{q}{c}\overrightarrow{A}\right)\Psi+\Psi\left(\hat{P}-\frac{q}{c}\overrightarrow{A}\right)\Psi^*\right) 
\end{equation}
taking into consideration eq. (\ref{eq:3}) we finally get that

$$\overrightarrow{\xi}_\lambda=-\frac{\hbar}{2M}\frac{\overrightarrow{\nabla}\rho}{\rho}+i\frac{q}{Mc}\overrightarrow{A}$$
We understand from the above equation that 

\begin{equation} \label{eq:8}
Re\overrightarrow{\xi}_\lambda=-\frac{\hbar}{2M}\frac{\overrightarrow{\nabla}\rho}{\rho}
\end{equation}

and that 
\begin{equation} \label{eq:9}
Im\overrightarrow{\xi}_\lambda=\frac{q}{Mc}\overrightarrow{A}
\end{equation}
When it comes to the current velocity we get
$$\overrightarrow{\eta}_\lambda=\frac{1}{2M\rho} (\Psi^*\left(\hat{P}-\frac{q}{c}\overrightarrow{A}\right)\Psi-\left(\Psi\left(\hat{P}-\frac{q}{c}\overrightarrow{A}\right)\Psi^*\right)$$
$$\overrightarrow{\eta}_\lambda=\frac{1}{2M\rho}\left(\Psi^*\hat{P}\Psi-\Psi\hat{P}\Psi^*\right)$$ 

Now let’s be more specific and let’s consider a system in which we have an infinitely long cylinder of radius $a$. Inside the cylinder there is constant magnetic field $\overrightarrow{B}=B\overrightarrow{e}_z$. Outside the cylinder the vector potential will be $\overrightarrow{A}=\frac{B\alpha^2}{2r}\overrightarrow{e}_\theta=\frac{\Phi_B}{2r\pi}\overrightarrow{e}_\theta $ (where $\Phi_B $ is the magnetic flux), while the magnetic field is zero.
The change in momentum due to the vector potential causes a transformation in Schrödinger’s equation 
$$\left(\overrightarrow{\nabla}_{(\lambda)}^2+k^2\right)\Psi=0$$
where we have that $\overrightarrow{\nabla}_{(\lambda)}=\overrightarrow{\nabla}+i\frac{\lambda}{r}\overrightarrow{e}_\theta$ and  $\lambda=-\frac{qBa^2}{2{\hbar}c}$. Also let us assume that the particle must be restricted in a region between two cylinders. The one is of radius  $r=\alpha$ and the other of radius $r=b>\alpha$. The solution of the above equation (taking into account that $\Psi(\alpha)=\Psi(b)=0$) will be
\begin{equation} \label{eq:10} 
\Psi(r,\theta)=N_{v,n}J_v\left(\frac{\tau_{v,n}}{d}(r-\alpha)\right)e^{im\theta} 
\end{equation}
where $\tau_{v,n}$ is the $n-$th root of the $v-$Bessel function, while also $d=b-a$ and $v=|m+\lambda|$. Using eqs. (\ref{eq:2}) and (\ref{eq:10}) we get
\begin{equation} \label{eq:11}
\overrightarrow{\eta}_\lambda=\overrightarrow{\eta}_m=\frac{m{\hbar}}{Mr}\overrightarrow{e}_\theta
\end{equation}
Now we define the quasi-current velocity and quasi-dispersive velocity from the equations $\overrightarrow{v}_\lambda=\frac{\overrightarrow{\Gamma}_\lambda}{\rho}$ and $\overrightarrow{w}_\lambda=\frac{\overrightarrow{\Delta}_\lambda}{\rho}$. For these quantities, using the eqs. (\ref{eq:5}),(\ref{eq:6}),(\ref{eq:8}),(\ref{eq:9}) and (\ref{eq:11}) we obtain 

\begin{equation} \label{eq:12}
\overrightarrow{v}_\lambda=\overrightarrow{\eta}_m+Im\left(-\overrightarrow{\xi}_\lambda\right)
\end{equation}
and

\begin{equation} \label{eq:13}
\overrightarrow{w}_\lambda=Re\overrightarrow{\xi}_\lambda 
\end{equation}

Also we notice that, if we multiply both sides of eq. (\ref{eq:12}) with the probability density, we get 
$$\overrightarrow{\Gamma}_\lambda=\overrightarrow{J}_\lambda+Im\left(-\overrightarrow{D}_\lambda\right)$$ or in a different form
\begin{equation} \label{eq:14}
\overrightarrow{\Gamma}_\lambda=\frac{i\hbar}{2M}\left(\Psi\overrightarrow{\nabla}\Psi^*-\Psi^*\overrightarrow{\nabla}\Psi\right)-\frac{q}{Mc}\overrightarrow{A}\rho
\end{equation}
 The last equation we have written down has been derived also by Aharonov and Bohm \cite{ahaBohm1959} from the continuity equation for time-dependent states.

\section{Comparison with other quantum systems} 
In this chapter, we will examine our results in the context of other quantum systems. Eq. (\ref{eq:12}) (reveals that the quasi-current velocity is composed of a current velocity, which is quantized and remains constant independent of the vector potential, and a velocity due to diffusion. It could be argued that the diffusion velocity we refer to is, in fact, the velocity the particle acquires as a result of its interaction with an external field, such as a quantum field (in this case).

The presence of an $Im$ and a minus sign may cause concern for some readers. However, we will see later that diffusion is typically a result of the interaction of the particle's quantum field with an external field via a force. In our system, we did not use any force, so this type of diffusion is acceptable. It is not obtained using the traditional method of solving the Schrödinger equation that contains a scalar potential. Regarding the sign, we can explain it based on the definition of diffusion. Suppose we have an electron moving in the region between the two cylinders. Due to the presence of the vector potential, we will have an additional angular momentum to the left. However, $Im\overrightarrow{\xi}_\lambda$ points to the right. This is because, by definition, the direction of the diffusion vector is opposite to the direction of the largest fluid density. This is the reason for calling $Im\left(-\overrightarrow{\xi}_\lambda\right)$ the velocity due to diffusion, while $\overrightarrow{\xi}_\lambda$ is the dispersive velocity. Alternatively, however, if we wanted to get rid of the minus sign, we could express eq. (\ref{eq:12}) in the following form:
$$\overrightarrow{v}_\lambda=\overrightarrow{\eta}_m+Im\overrightarrow{\zeta}_\lambda$$

In general, the osmotic velocity has the opposite role to that of diffusion. It pushes the particle to higher concentration areas. 

We have previously mentioned that the quasi-current velocity contains a diffusion term, which may seem unusual. However, this is not an isolated case. There are systems, such as the Gaussian wave packet, where we can observe that the current velocity contains a diffusion term. The probability density of a Gaussian wave packet is given by:
$$\rho_{g}(x,t)=\sqrt{\frac{2}{\pi}}\frac{1}{\epsilon}e^{-2(x-u_{0}t)^2/\epsilon^2}$$ where $\epsilon(t)=\alpha\sqrt{1+\left(\frac{4\hbar^2t^2}{m^2\alpha^4}\right)}$.

According to \cite{mita2003dispersive} we have 
$$\overrightarrow{J}_g(x,t)=\rho_g(x,t)u_0\overrightarrow{e}_x+\frac{mt}{T}\overrightarrow{D}_g(x,t)$$ 
written in an equivalent way
$$\overrightarrow{\eta}_g(x,t)=u_0\overrightarrow{e}_x+\frac{mt}{T}\overrightarrow{\xi}_g(x,t)=u_0\overrightarrow{e}_x+\frac{2t\hbar}{\epsilon^2T}(x-u_0t)\overrightarrow{e}_x$$ where $T=\frac{m\alpha^2}{2\hbar}$

 The first term on the right-hand side of this equation corresponds to the center-of-mass motion of the Gaussian wave packet, which has a constant velocity $u_0$. The second term is a diffusion term that arises due to the interaction of our quantum field with a field similar to that of the harmonic oscillator (in a zero-point energy state)  before $t=0$, which occurs via an inhomogeneous restraining force which has the form $\overrightarrow{F}=-kx\overrightarrow{e}_x$ (throughout the length of this paper we will refer to this field as the harmonic oscillator field). If we remove this external force at $t=0$, we will observe diffusion of the fluid. The motion of the particle, which is carried by the fluid, is determined by an inhomogeneous quantum force (this physical variable was introduced by Bohm\cite{bohm1952suggested}, but in our description it represents the force exerted by the quantum fluid to the particle), whose direction is opposite to that of the classical external force. This force is given by the equation:
 $$\overrightarrow{F}_{Q,g}(x,t)=\frac{\hbar^2}{2m}\overrightarrow{\nabla}\left(\frac{\nabla^2\sqrt{\rho_g(x,t)}}{\sqrt{\rho_g(x,t)}}\right)\Rightarrow$$
 $$\overrightarrow{F}_{Q,g}(x,t)=\frac{4\hbar^2}{m\epsilon^4}(x-<x>_t)\overrightarrow{e}_x$$
 
 The quantum force becomes stronger as we move away from the center of the wave packet, which is intuitive since the diffusion of the wave packet becomes more significant as we move farther away from its center. This is due to the fact that before $t<0$, the restricting force was stronger for greater $x$ values, which led to the opposite result after its abolishment. We can think of this in terms of fluid pressure, where if the pressure is exerted on the left, and we suddenly remove it, we observe a sudden expansion towards the right.

This observation provides evidence for the interaction of our quantum field with another field. However, we will delve into this topic in more detail in Chapter 5.

The Airy wave packets are a particularly interesting type of wave packet that exhibit self-acceleration in the absence of an external force field and do not disperse. These wave packets are given by the initial state,
$$\Psi_A(x,0)=Ai\left(\frac{(2mk)^{1/3}}{\hbar^{2/3}}x\right)$$
where $Ai$ is the Airy function. 
According to the work of Berry and Balazs \cite{berry1979nonspreading}, the time-evolved state of this wave packet is given by:
$$\Psi_A(x,t)=Ai\left[\frac{(2mk)^{1/3}}{\hbar^{2/3}}\left(x-\frac{kt^2}{2m}\right)\right]e^{i\frac{kt}{\hbar}\left(x-\frac{kt^2}{3m}\right)}$$
This expression shows that the Airy wave packet maintains its shape and does not disperse as time evolves, while also undergoing self-acceleration due to the quadratic term in the position.

Although we will not delve extensively into this topic, it is worth noting that the peculiar behavior of the Airy wave packets can also be understood in terms of the external force field that created our initial state ($\overrightarrow{F}_{ext}=-k\overrightarrow{e}_x$, for $E=0 $). The reason behind the non-dispersive nature of the wave packet is the homogeneity of the quantum force generated by the quantum field (due to its initial interaction with the external force field). The force acts on the particle in the positive $x$-axis direction, causing the wave packet to accelerate to the right. The time-dependent quantum force can be expressed as:
$$\overrightarrow{F}_{Q,A}(x,t)=k\overrightarrow{e}_x$$
and is opposite in direction to the external force.

Previous interactions of our fluid with other fields can have significant effects on the new system where the fluid is introduced. This is also observed in the case of the Aharonov-Bohm effect, but we will discuss that in more detail in chapter 5.

We previously mentioned that the quasi-probability current and quasi-diffusion current have properties similar to their traditional counterparts. In multi-dimensional systems, we observe that the quasi-probability current and quasi-diffusion current are orthogonal, as expressed by the following equation:
$$\overrightarrow{\Gamma}_\lambda\cdot\overrightarrow{\Delta}_\lambda=0$$
This property is analogous to the orthogonality between the probability current and diffusion current in probability theory. It is a useful property that allows us to better understand and analyze the behavior of the wave packet in multi-dimensional systems.

Similarly, in the absence of a magnetic field, the traditional probability and diffusion current are also orthogonal, given by
$$\overrightarrow{J}_{\lambda=0}\cdot\overrightarrow{D}_{\lambda=0}=0$$

This orthogonality also holds true for other systems, such as the hydrogen atom, where the wave function takes the form:
 $$\Psi_{n,l,m_l}(r,\theta,\phi)=R_{n,l}(r)\Theta_l^m(\theta)e^{im\phi}$$ 
 According to \cite{mita2003dispersive} we have 
 $$\overrightarrow{D}_{n,l,m_l}(r,\theta)=-\frac{\hbar}{4{\pi}m_e}\left[\overrightarrow{e}_r\frac{d}{dr}\left(R_{n,l}(r)\right)^2\left(\Theta_l^m(\theta)\right)^2+\overrightarrow{e}_\theta\frac{1}{r}\left(R_{n,l}(r)\right)^2\frac{d}{d\theta}\left(\Theta_l^m(\theta)\right)^2\right]$$and
 $$\overrightarrow{J}_{n,l,m_l}(r,\theta)=\overrightarrow{e}_\phi\frac{m_l\hbar}{Mrsin\theta}\rho_{n,l,m_l}(r)$$Therefore we get
 $$\overrightarrow{J}_{n,l,m_l}(r,\theta)\cdot\overrightarrow{D}_{n,l,m_l}(r,\theta)=0$$

\section{Energy and angular momentum in the Aharonov-Bohm effect}
In this chapter, we will employ the principles of quantum hydrodynamics and diffusion processes to calculate the angular momentum and energy of our system. By incorporating the insights from eq. (\ref{eq:4}), we can derive a precise expression for the same.

 $$ \Psi^*\left(\hat{r}\times\hat{P}'\right)\Psi=M(\hat{r}\times\overrightarrow{\Gamma}_\lambda+i\hat{r}\times\overrightarrow{\Delta}_\lambda) $$
 Since $M\rho\hat{r}\times Re\overrightarrow{\xi}_\lambda=0$
 $$\Psi^*\left(\hat{r}\times\hat{P}'\right)\Psi\overrightarrow{e}_z=M\rho\left(\hat{r}\times\overrightarrow{\eta}_m\right)\overrightarrow{e}_z+M\rho\left(\hat{r}\times Im\left(-\overrightarrow{\xi}_\lambda\right)\right)\overrightarrow{e}_z\Rightarrow$$
 $$ \Psi^*\hat{L}_{z,tot}\Psi=\Psi^*\hat{L}_{z}^{\left(\zeta\right)}\Psi+\Psi^*\hat{L}_{z}^{\left(\eta\right)}\Psi$$
 
Therefore, by utilizing the concepts of quantum hydrodynamics and diffusion processes, we can accurately determine the angular momentum and energy of our system. By considering eq. (\ref{eq:4}), we can obtain the total angular momentum density by adding the contributions from both the current velocity and the diffusion processes. Integrating both sides of the equation enables us to obtain the total angular momentum of the system, as
 
 $$\left<\hat{L}_{z,tot}\right> = \hbar(m+\lambda)$$
 we also know that:

$$\left<\hat{L}_{z}^{(\zeta)} \right> = \hbar\lambda$$

 It is evident that the angular momentum arising from this particular type of motion is not subject to quantization, unlike the canonical angular momentum that emerges from the current velocity. This leads us to infer that the lack of quantization in the additional angular momentum can be traced back to its origin, which is a distinct form of quantum velocity, namely, the velocity induced by diffusion, which is the imaginary part of the osmotic velocity.
 
Next, let us turn our attention to the energy of the system. To begin with, we need to determine the energy density, which can be represented as follows:
$$\epsilon_{k,(\lambda)}=\Psi^*\frac{\left(\hat{P}-\frac{q}{c}\overrightarrow{A}\right)^2}{2M}\Psi$$or
$$\epsilon_{k,(\lambda)}=-\frac{\hbar^2}{2M}\Psi^*\overrightarrow{\nabla}_{(\lambda)}^2\Psi=-\frac{\hbar^2}{2M}\Psi^*\left(\overrightarrow{\nabla}+i\frac{\lambda}{r}\overrightarrow{e}_\theta\right)^2\Psi\Rightarrow$$
$$\epsilon_{k,(\lambda)}=-\frac{\hbar^2}{2M}\Psi^*\overrightarrow{\nabla}^2\Psi+T_{m,\lambda}(r)$$ where
 $T_{m,\lambda}(r)=\frac{\lambda^2\hbar^2}{2Mr^2}\rho+\frac{m\lambda\hbar^2}{Mr^2}\rho$

The initial component corresponds to the kinetic energy density associated with a wave function $\Psi$, which is identical to the expression provided in eq. (\ref{eq:10}). So, according to \cite{mita2021schrodinger}
 $$-\frac{\hbar^2}{2M}\Psi^*\overrightarrow{\nabla}^2\Psi=\epsilon_{k,\Sigma}=\frac{1}{2}M\rho\left(\overrightarrow{\eta}_\Sigma^2+\overrightarrow{\xi}_\Sigma^2\right)$$
 however, it is also worth noting that:
 $$\overrightarrow{\eta}_\Sigma=\overrightarrow{\eta}_m $$
 and
 $$\overrightarrow{\xi}_\Sigma=Re\overrightarrow{\xi}_\lambda$$
 Therefore we get
 $$\epsilon_{k,(\lambda)}=\frac{1}{2}M\rho\left(\left(\overrightarrow{\eta}_m+Im\left(-\overrightarrow{\xi}_\lambda\right)\right)^2+\left(Re\overrightarrow{\xi}_\lambda\right)^2\right)$$
 or
 $$\epsilon_{k,(\lambda)}=\frac{1}{2}M\rho\left(\overrightarrow{v}_\lambda^2+\overrightarrow{w}_\lambda^2\right)=\frac{1}{2}M\frac{\overrightarrow{\Gamma}_\lambda^2+\overrightarrow{\Delta}_\lambda^2}{\rho}$$
 Alternatively, we can express this observation in a different manner:
 $$\epsilon_{k,(\lambda)}=\Psi^*\frac{\hat{L}_{z,tot}^2}{2Mr^2}\Psi+\frac{1}{2}M\rho\left(Re\overrightarrow{\xi}_\lambda\right)^2$$

We can observe that in a system with a vector potential, the total energy density can be decomposed into two distinct components: the energy density arising from the oscillatory motion of the particle in the fluctuating field (or fluid) along the r-axis, and the energy density originating from the rotational motion. The former is a result of quasi-diffusion, while the latter is due to quasi-current velocity. It is worth noting that this behavior is not unique to our system, as we can observe similar phenomena in other systems, such as the hydrogen atom, where traditional diffusion and probability current contribute to the energy density as well in the same way \cite{mita2021schrodinger}.

\section{The role of forces in the Aharonov-Bohm effect}

In our theory, and according to Bohm, the osmotic velocity (or the dispersive velocity) is a variable that can only be obtained through the interaction of the quantum field of our particle with another field, mediated by a force \cite{bohm1989non}. This force requires the presence of a scalar potential. For instance, in the case of stationary states in one dimension, the quantum Euler equation \cite{mita2021schrodinger} applies, i.e.
 $$\xi\frac{d\xi}{dx}-\beta^2\frac{d^2\xi}{dx^2}-\frac{1}{m}\frac{dV}{dx}=0$$
 
In systems like these, the value of $\eta$ is zero. This can be problematic in Bohmian mechanics because it would suggest that the particle does not move. However, current velocity may be non-zero in a system of more than one dimensions. In such cases, we have
 $$\frac{\partial\overrightarrow{\eta}}{\partial t}+\frac{1}{2}\overrightarrow{\nabla}\left(\eta^2\right)=\frac{1}{2}\overrightarrow{\nabla}\left(\xi^2\right)-\beta^2\nabla^2\overrightarrow{\xi}-\frac{1}{m}\overrightarrow{\nabla}V$$
 The equation in question incorporates both the current velocity and the scalar potential, but in reality, these two variables are not necessarily correlated with each other, especially in bound states. For example, in the case of the hydrogen atom $(V(r)=-e^2/4\pi\epsilon_{0}r)$ we get
 $$\overrightarrow{\eta}_{n,l,m_l}(r,\theta)=\overrightarrow{e}_\phi\frac{m_l\hbar}{Mrsin\theta}$$ a variable independent of the charge and $\epsilon_0$, as it depends only on the quantization of states (it just depends on $m_l$, which is the magnetic quantum number). The osmotic velocity, however, is dependent on the scalar potential and this fact arises from the osmotic velocity's dependence on the charge and $\epsilon_0$. Therefore, it is important to recognize that this additional current velocity we obtain in our system arises from diffusion in order to justify a field interaction.
 
 At the outset of this chapter, our goal is to derive the quantum force for our system. The quantum potential, which was first introduced in Bohmian mechanics, can be obtained via the transformed Hamilton-Jacobi equation, as demonstrated by Philippidis et al. \cite{philippidis1982aharonov}.
 $$\frac{\partial S} {\partial t}+\frac{1} {2m}\left(\overrightarrow{\nabla}S-\frac{q} {c}\overrightarrow{A}\right)^2+V+Q=0$$
 where $Q$ is the quantum potential. The above equation holds for a wavefunction of the form $\Psi=Re^{iS/\hbar}$.
 
 Hence we find that
 \begin{equation} \label{eq:15}
 Q_\lambda=-\frac{1}{2}M\left(\overrightarrow{\eta}_m+Im\left(-\overrightarrow{\xi}_\lambda\right)\right)^2=-\frac{1}{2}M\overrightarrow{v}_\lambda^2=-\frac{\hbar^2\left(m+\lambda\right)^2}{2Mr^2}\Rightarrow
 \end{equation}
 $$\overrightarrow{F}_{Q,\lambda}=-\overrightarrow{\nabla}Q_\lambda=-\frac{\hbar^2\left(m+\lambda\right)^2}{Mr^3}\overrightarrow{e}_r=-M\frac{\overrightarrow{v}_\lambda^2}{r}\overrightarrow{e}_r$$
 This is a logical outcome because the quantum force is responsible for determining the trajectory of the particle. Therefore, in the case of a rotational orbit, a centripetal force arises naturally.
 
Next, we will explore some aspects of the magnetic force. Typically, this force is attributed to the presence of a magnetic field. However, we will attempt to offer a different interpretation by re-examining our understanding of magnetic fields.

According to \cite{arbab2011analogy}, the motion of a vortex in a rotational flow is described by the following force:
$$\overrightarrow{F}_m=-m\overrightarrow{v}\times \left(\overrightarrow{\nabla}\times \overrightarrow{v}\right)$$
In this case $m$ is the fluid’s mass. 
If we divide both sides of the equation by the volume of the fluid, we obtain
\begin{equation} \label{eq:16}
\overrightarrow{f}_m=\frac{\overrightarrow{F}_m}{V_{volume}}=-\rho_m\overrightarrow{v}\times \left(\overrightarrow{\nabla}\times \overrightarrow{v}\right)=-\rho_m\overrightarrow{v}\times \overrightarrow{\omega}
\end{equation}
where $\rho_m$ is the fluid density and $\overrightarrow{\omega}$ the vorticity field.

The result is the force per unit volume that characterizes the motion of fluid density. Now, let's consider the Lorentz force, which can be written as:
\begin{equation} \label{eq:17}
\overrightarrow{F}_B=\frac{q}{c}\overrightarrow{v}_f\times \overrightarrow{B}=\frac{q}{c}\left(\overrightarrow{v}_i+\delta\overrightarrow{v}\right)\times \overrightarrow{B}
\end{equation}
In the context of our quantum system, which consists of a quantum fluid inside a cylinder, we can express this force as follows:
$$\overrightarrow{F}_B=\frac{q}{c}\left(\overrightarrow{\eta}_{in}+Im\left(-\overrightarrow{\xi}_{in}\right)\right)\times \overrightarrow{B}$$
Given that $\overrightarrow{B}=\overrightarrow{\nabla}\times \overrightarrow{A}$, $\delta\overrightarrow{v}=Im\left(-\overrightarrow{\xi}_{in}\right)=-\frac{q}{Mc}\overrightarrow
{A}$ and $$\overrightarrow{\nabla}\times \overrightarrow{\eta}_{in}=\beta^2\overrightarrow{\nabla}\times \left(\overrightarrow{\nabla}(2S_{in})\right)=0$$ we find the force
$$\overrightarrow{F}_B=-M\overrightarrow{v}_f\times \left(\overrightarrow{\nabla}\times \overrightarrow{v}_f\right)=-M\overrightarrow{v}_f\times \overrightarrow{\omega}_{quantum,in}$$
where $\overrightarrow{\omega}_{quantum,in}=-\frac{q}{Mc}\overrightarrow{B}=\overrightarrow{\nabla}\times Im\left(-\overrightarrow{\xi}_{in}\right)\neq 0$

The magnetic force is equivalent to the force described in eq. (\ref{eq:16}). When a particle is in a region with a magnetic field, it behaves as if it were in a rotating vortex with a vorticity of $\omega_{quantum,in}$. A similar equation was also introduced by Takabayashi \cite{takabayasi1}. 

The magnetic field serves as a tool for calculating this vorticity field, which is generated by the interaction between the field and the current. Similarly, the vector potential is used to calculate the velocity due to diffusion. Our goal is to describe velocities as more fundamental quantities than magnetic fields and vector potentials. To achieve this, we will demonstrate that the seemingly peculiar magnetic force, which is said to arise due to the presence of a magnetic field, is actually a well-known Newtonian force in classical hydrodynamics. Thus, it is not the force or the magnetic field that is strange, but rather the type of interaction between the quantum fields that leads to the existence of this force. We will refer to this in more detail later.

For now, having in our mind the vorticity, let’s find some more similarities with classical hydrodynamics. So we have that
$$Im\overrightarrow{\xi}_{\lambda,out}=\frac{q} {Mc}\frac{Ba^2} {2r}\overrightarrow{e}_\theta=>Im\left(-\overrightarrow{\xi}_{\lambda,out}\right)=\frac{\omega_{quantum,in}\alpha^2}{2r}\overrightarrow{e}_\theta$$ 
and for a fluid outside
$$Im\overrightarrow{\xi}_{in}=\frac{q} {Mc}\frac{Br} {2}\overrightarrow{e}_\theta=>Im\left(-\overrightarrow{\xi}_{in}\right)=\frac{\omega_{quantum,in}r}{2}\overrightarrow{e}_\theta$$

In classical hydrodynamics, two velocities similar to those in our quantum system are derived in \cite{green2012fluid}. These velocities are known as velocity in case of non-rotational vortex and velocity in case of rotational vortex. It's worth noting that a particle (which is carried by the quantum fluid) outside the magnetic field region has behavior of an irrotational vortex, while inside the behavior of a rotational vortex. Furthermore, the velocity due to diffusion increases as we move closer to the current, indicating the direct role of the electric current in causing a rotational diffusion of the particle.

Let's explore further similarities between classical hydrodynamics and our quantum system. Specifically, let's consider the quantum potential for a particle outside the cylinder, which we have already calculated (eq. (\ref{eq:15})). According to \cite{heifetz2015toward}, the quantum potential is a quantity equivalent to the pressure applied in a fluid. Moving on to the pressure for an irrotational vortex, according to \cite{green2012fluid}, we can express it using the equation:

$$P(r,z)=-\frac{\rho\Gamma^2}{2{\pi}r^2}=-\frac{1} {2}\rho\overrightarrow{u}_\theta^2$$

where $u_\theta =\frac{\Gamma}{2{\pi}r}\overrightarrow{e}_\theta$ is the flow velocity of the vortex and $\Gamma$ the vorticity flux.
We can observe that the pressure for an irrotational vortex is equivalent to the quantum potential we derived earlier in eq.(\ref{eq:15}). Therefore, we can identify the motion of the quantum particle with the motion of an irrotational vortex.

There are two methods for describing the magnetic force. The first involves the use of the equation $\overrightarrow{F}_1=\frac{q}{c}\overrightarrow{v}_f\times \overrightarrow{B}$ while the second employs the equation $\overrightarrow{F}_2=-M\overrightarrow{v}_f\times \left(\overrightarrow{\nabla}\times \overrightarrow{v}_f\right)$. Having this in mind, there are two ways to interpret the presentation of magnetic fields ($B$). The first is that they are the source of magnetic forces. The second way is to say that $\overrightarrow{\omega}=\overrightarrow{\nabla}\times \overrightarrow{v}_f=-\frac{q}{Mc}\overrightarrow{B}=-\frac{q}{Mc}B\overrightarrow{e}_z=-\frac{q}{Mc}\left(\mu_0nI\right)\overrightarrow{e}_z$. In this case, $B$ is an empirical constant that includes $\mu_0$ (a constant), as well as variables such as $n$ and $I$ that are related to the surroundings of the magnetic field area (where $n$ represents the number of coils per unit length). The second interpretation helps us understand that the current is responsible for the appearance of the rotational velocity due to diffusion, and not some sort of magical force that exists inside the cylinder (i.e. the magnetic field). We can extend this way of thinking to a fluid outside the magnetic field area and explain why $Im\overrightarrow{\xi}_{\lambda,out}=\frac{q} {Mc}\frac{Ba^2} {2r}\overrightarrow{e}_\theta=\frac{q} {Mc}\frac{\left(\mu_0nI\right)\alpha^2}{2r}\overrightarrow{e}_\theta.$

This response effectively addresses the perspective that asserts the magnetic field or magnetic flux within a cylinder induces non-local changes in states, such as angular momentum or energy, for a particle located outside the magnetic field region. While this phenomenon may appear paradoxical and non-local to some, it can be comprehended when considering the interaction between the quantum field and the current. The current, playing a fundamental role in this scenario, provides a basis for further elucidation, which we can explore in greater detail.

In this sense, we can consider $B$ as a vorticity coefficient. This is why $B$ is non-zero inside and zero outside the magnetic field area, as inside the cylinder, we have vorticity, whereas outside, we do not. Therefore, when it comes to the outside area, it does not make sense to talk about a vorticity coefficient (and thus about a magnetic field).

One might question why, if we have this field interaction, we do not have a force that causes changes in our physical effects in the magnetic field-free area. The answer to this question is that, classically speaking, there may be no force that can cause an irrotational vortex. Even if such a force exists, someone could suggest that, at least in our particular quantum system, this force has the form of:
$$\overrightarrow{F}_{ext}=\frac{{\hbar^2}m}{Mr^3}(m+2\lambda)\overrightarrow{e}_r$$
or in an equaivalent form
\begin{equation} \label{eq:18}
\overrightarrow{F}_{ext}=\left(\frac{M{\delta}\overrightarrow{v}^2}{r}+\frac{M{\delta}\overrightarrow{v}\cdot\overrightarrow{v}_i}{r}\right)\overrightarrow{e}_r
\end{equation}
where $\delta\overrightarrow{v}=Im\left(-\overrightarrow{\xi}_{out}\right)$ and $\overrightarrow{v}_i=\overrightarrow{\eta}_m$

Let us consider two quantum systems, $A$ and $B$, where $A$ has a vector potential (our previous system) and $B$ (a new system) that has a scalar potential $V_{ext}=(\hbar^2 m)(m+2\lambda)/(2Mr^2)$, but no vector potential. Despite the absence of a vector potential in $B$, the two systems are identical since they produce the same states and energies.  It may seem that there is a physical difference between the two systems, as in $A$ we have a change in momentum and angular momentum while in $B$ these two variables remain unchanged ($\left<P\right> = 0$, $\left<L _z \right>={\hbar}m$). However, the motion of the particle is governed by the same quantum force, which is obtained from the Hamilton-Jacobi equation in both systems. The reason for the different values of momentum and angular momentum is simply due to the way the systems are presented. In system $A$, the transformation is performed to obtain the result of the interaction between the two quantum fields (extra diffusion, momentum, angular momentum, etc.). On the other hand, in system $B$, the momentum of the charge is considered as the canonical momentum before the interaction with the current, and the result is seen after the interaction between the two fields through the exertion of an external force (the result is the same to that of system $A$). Despite the different values of momentum and angular momentum in the two systems, all other physical quantities remain the same. Therefore, the difference is purely a matter of presentation, and there is no difference in physical meaning between the two systems.

Using system B, we can gain insight into the origin of the extra rotational diffusion observed in system A. As previously mentioned, a force field is required to obtain a diffusion term. Examining the external force more closely, we find that it is responsible for the acquisition of an extra quantum force, denoted as $\delta \overrightarrow{F}_{Q,\lambda}$. This force is given by $\delta \overrightarrow{F}_{Q,\lambda}=-\left(\frac{M{\delta}\overrightarrow{v}^2}{r}+\frac{M{\delta}\overrightarrow{v}\cdot\overrightarrow{v}_i}{r}\right)\overrightarrow{e}_r$, and is entirely responsible for the extra rotation observed in system A, contributing to the total centripetal force. Changes in states actually result from this extra rotation (rotational diffusion can be regarded as the direct consequence of this quantum force), but we will discuss this further later. Therefore, it is the rotational diffusion, rather than the change in states, that is the actual result of the force field.

Now let's examine eqs. (\ref{eq:18}) and (\ref{eq:17}) (which describes the magnetic force). Although these equations represent different forces, it's reasonable to expect that there are some similarities between them. In fact, we can see that both equations depend on both the current velocity $\overrightarrow{v}_i$ and the rotational velocity due to diffusion, which is $\delta\overrightarrow{v}$.

We can say that the differences in geometry between the two systems (inside and outside) result in a different force that describes the motion of the particle after the interaction. It is important to note that a force is a result of a specific interaction and not the cause. To illustrate this point, consider the centripetal force experienced when moving in a circle. This force exists because of the tangential velocity, but it is acquired through interactions with the ground, such as friction and slip coefficients. While we could express the slip coefficient as a field, we typically treat it as a coefficient that affects motion. Similarly, we do the same here by noting that $\overrightarrow{\omega}_{quantum,in}=-(qB/{Mc})\overrightarrow{e}_z =f(B)\overrightarrow{e}_z$, where $B$ is the magnetic field strength.

The importance of the rotational diffusion term lies in its ability to explain the change in probability densities and energies of a quantum system. According to Bohm, a diffusion term arises from the interaction of the fluid with a force field, which ,in this system, can be produced by the scalar potential in system B. In contrast to the mathematical wave approach, where the vector potential affects the movement of the particle, our physical field pre-existed at $r\geq a$ before the placement of the potential walls. The probability density is related to a part of this physical fluctuating field, and the velocity due to diffusion is a quantum quantity independent of the probability density, which means that the fluid existed in those areas where the probability density is zero. Furthermore, we know that the state of our fluid in a particular field may depend on previous field interactions. This was observed in the case of Airy wave packets or Gaussian wave packets (where the produced diffusion in a free-field area defined the time-evolution of the probability density). Therefore, understanding the origin and importance of the diffusion term can provide insight into the behavior of our present quantum systems. Besides, that's a fundamental property of Markov processes that stimulate quantum dynamics (see Nelson's paper \cite{nelson1966derivation} for more details). A present state always depends on a past state.

We acknowledge the significance of our previous discussion on quantum forces, as demonstrated by Becker and other researchers who utilized quantum forces to explain the phenomenon (\cite{becker2019asymmetry},\cite{becker2017observation}). Notably, quantum force exhibits proximity to electric currents and assumes non-zero values in regions where the probability density is zero. This observation reinforces our earlier assertion regarding its influence on current interactions, as well as its connection to the diffusion term within our system. It is worth noting that osmotic velocity shares similarities with the quantum force due to their association with the stochastic motion of particles and the participation of both quantities in the quantum Euler equation as the stochastic terms. However, we diverge from the same explanatory path regarding this effect, despite recognizing the utility of quantum forces in comprehending the behavior and temporal evolution of wave packets during our previous discussion.

On the other hand, of course, having considered the role of quantum forces in the time evolution of states even in free field areas, someone could extend his thoughts in our bound system. So. it can be proposed that despite the abolishment of the external force due to the interaction with the current, by confining an amount of fluid, the quantum force seemed to exist even during the present state of our fluid, resulting in the rotational diffusion. Anyway, we do not intend to deepen further into that topic.

\section{The physical importance of the quantum background field}

The fields can be treated as real physical fields that can interact with each other, and not just as mathematical constructions, due to the Newtonian behavior of the body in the quantum fluid, which is identical to the behavior of a body in a classical fluid. The quantum diffusion coefficient is inversely proportional to the mass, as in classical fluid mechanics, which means that smaller masses are more susceptible to the motion of the fluid and can be guided by it more easily. This is why small masses exhibit more quantum mechanical or fluid-like behavior, resulting in non-classical trajectories. Therefore, the physical fields in quantum mechanics are not just mathematical constructs, but they have a physical effect on the behavior of particles, as observed in classical fluids.

Combining this knowledge with the Gaussian wave packet and its diffusion, we can observe a previous interaction or pressure of the whole fluctuating field with the field of the harmonic oscillator. This diffusion of the field is related entirely to the field itself and not the point-like mass. The diffusion of the physical fluid behaves always regardless of mass, but the diffusion of the particle within the fluid or the diffusion of the mathematical fluid known as probability density is related to the mass of the particle. For very large masses, the diffusion of the particle inside the fluid, or the diffusion of the probability density, will be very small because the mass is less responsive to the motion of the fluid that carries it. This is consistent with classical mechanics, where a larger mass requires a greater force to move it, and the same holds true for the diffusion of the particle in a fluid. Therefore, the mass of a particle affects its diffusion within a physical fluid, which also applies to the mathematical fluid of the probability density in quantum mechanics.

It is important to note that magnetic and electric fields are not simply mathematical constructs, but they are actually existent due to the existence of the quantum field. That is,in fact, the deepest reason why we claim that the motion of the charge creates the magnetic or electric field, which in turn affects the motion of other charges. 

In our overall description, we adopt a quantum framework that encompasses both the excluded charge outside the magnetic field area and the electric current as quantum particles. Notably, Vaidman (\cite{vaidman2012role}, \cite{vaidman2015reply}) has also provided a quantum description of this phenomenon. However, his interpretation differs from ours as he viewed it through the lens of quantum entanglement, despite considering the source of potentials as quantum particles.

\section{Time-dependent states in the Aharonov-Bohm effect}

In this chapter, we will discuss the concept of time-dependent states. It is important to acknowledge that the justification for exploring the interaction between the quantum field of charge and the quantum field of current lies in the rotational diffusion we observe. This additional rotational velocity directly contributes to the alteration of states and leads to a different oscillatory behavior along the $r$-axis. Therefore, it is not exactly the state changes themselves that warrant our consideration of the interaction between our quantum field and another field, but rather the diffusion phenomenon, which we know is always a consequence of such an interaction (in contrast to the current velocity) and which, as we know, has the property of defining probability distributions (we will study this further later). Furthermore, when addressing time-dependent states, we encounter the following transformation,
$$\Psi_B(\overrightarrow{r},t)=e^{iq\Lambda(\overrightarrow{r},t)/c}\Psi_{B=0}(\overrightarrow{r},t)$$ where $\overrightarrow{A}=\overrightarrow{\nabla}\Lambda$

In the present system we have that
$$Re\overrightarrow{\xi}_B(\overrightarrow{r},t)=-\frac{\hbar}{2M}\frac{\overrightarrow{\nabla}\rho_{B}(\overrightarrow{r},t)}{\rho_B(\overrightarrow{r},t)}$$
However, since it is evident that $\rho_{B}(\overrightarrow{r},t)=\rho_{B=0}(\overrightarrow{r},t)$ we get that

$$Re\overrightarrow{\xi}_B(\overrightarrow{r},t)=Re\overrightarrow{\xi}_{B=0}(\overrightarrow{r},t)$$ or $$Re\overrightarrow{\xi}_B(\overrightarrow{r},t)=\overrightarrow{\xi}_{B=0}(\overrightarrow{r},t)$$ Also,

$$Im\overrightarrow{\xi}_B(\overrightarrow{r},t)=\frac{q}{Mc}\overrightarrow{A}(\overrightarrow{r},t)=\frac{\hbar}{M}\overrightarrow{\nabla}\delta S(\overrightarrow{r},t)=\delta \overrightarrow{\eta}_B(\overrightarrow{r},t)$$

By examining the three preceding equations, it becomes evident that the real component of diffusion remains constant, as does the probability density. Therefore, the only means of justifying the interaction between quantum fields is through the imaginary component of diffusion. What sets this system apart from other "traditional" systems is the remarkable fact that diffusion in this context is intricately tied to the phase of the wave function—the additional phase obtained within the wave function.

Another way of writing the above equation is
\begin{equation} \label{eq:19}
\overrightarrow{\eta}_B(\overrightarrow{r},t)=\overrightarrow{\eta}_{B=0}(\overrightarrow{r},t)+Im\overrightarrow{\xi}_B(\overrightarrow{r},t)
\end{equation}
where $\overrightarrow{\eta}_{B=0}(\overrightarrow{r},t)=\beta^2\overrightarrow{\nabla}\left(2S^{(0)}(\overrightarrow{r},t)\right)$ and $S^{(0)}$ the phase that the wavefunction has regardless of the interaction.

Certainly, it is evident that in this scenario, the current velocity also includes a diffusion term. However, if we aim to develop a description that aligns more closely with our bound state system, we need to examine the behavior of the quasi-current velocity. Assuming that $\overrightarrow{v}_B(\overrightarrow{r},t)$ represents the quasi-current velocity, we can express the equation as follows:
$$\overrightarrow{v}_B(\overrightarrow{r},t)=\frac{1}{M\rho_B(\overrightarrow{r},t)}Re\left(\Psi_B^*(\overrightarrow{r},t)\hat{P}'\Psi_B(\overrightarrow{r},t)\right)$$
Due to the fact that the probability density and momentum density remain unaltered in the presence of the vector potential (which can be easily demonstrated) in time-dependent states, it follows that:
$$\overrightarrow{v}_B(\overrightarrow{r},t)=\overrightarrow{\eta}_{B=0}(\overrightarrow{r},t) $$ 
or in an equivalent form
 $$\overrightarrow{v}_B(\overrightarrow{r},t)=\overrightarrow{\eta}_B(\overrightarrow{r},t)+Im\left(-\overrightarrow{\xi}_B(\overrightarrow{r},t)\right)$$
The above equation is equivalent to the eq. (\ref{eq:12}) for a bound state system (where we call $\overrightarrow{\eta}_\lambda$ as $\overrightarrow{\eta}_m$, because it only depends on $m$ and not on $\lambda$.

Of course, if we multiply both sides of the above equation with the probability density, we obtain an expression for the quasi-probability current. This expression is equivalent to the expression for the probability current, which was presented by Aharonov and Bohm in their paper. The same expression also exists in bound states as seen from eq. (\ref{eq:14}).

\section{Hidden variables and the Aharonov-Bohm effect}

In this chapter, our focus lies on understanding the infinite physical effects that the transformation can have on the charge in time-dependent states (including bound states), primarily due to the potential infinity of $\Lambda$. In the presence of a magnetic field and current, it is important to note that the mean momentum, energy, and probability density remain unchanged in these time-dependent states. Nevertheless, we emphasize that changes occur in phase and fluid mechanical quantities, which hold significance for our analysis. While one might argue that vector potentials possess a greater fundamental nature than magnetic fields, we do not view magnetic fields and vector potentials as fundamental entities but rather as tools. In contrast, diffusion velocity, driven by the interaction between quantum fields, is the fundamental aspect we consider. Thus, instead of attributing the existence of numerous physical outcomes to different vector potentials, our focus is on explaining the occurrence of distinct diffusion velocities while acknowledging the unaltered external cause (the current) that gives rise to them through the interaction of quantum fields.

More specifically, we have that
$$Im\left(-\overrightarrow{\xi}_B^\Lambda\right)=-\frac{q}{Mc}\overrightarrow{A}' \Rightarrow$$ 
$$Im\left(-\overrightarrow{\xi}_B^\Lambda\right) =Im\left(-\overrightarrow{\xi}_B^0\right)-\frac{q}{Mc}\overrightarrow{\nabla}\Lambda$$
where the above equation holds for time-dependent states or the bound states that we studied earlier.
In this context, it is important to note that the osmotic velocity extracted from this system exhibits peculiar properties, as we have previously discussed. In both bound states and time-dependent states, this velocity contributes to the overall current velocity (although this phenomenon is observed sometimes in time-dependent states, and more specifically, in the Gaussian wave packets, it is considered as unusual in bound states), referred to as the quasi-current velocity. Additionally, in time-dependent states, the osmotic velocity is also associated with an additional phase factor, further complicating its behavior and impact.
Also we know that in bound states
$$E[\overrightarrow{\zeta}_\lambda]=\int{\rho\overrightarrow{\zeta}_\lambda{d\tau}}=\frac{\hbar}{2M}\int{\overrightarrow{\nabla}\rho{d\tau}}-\frac{iq}{Mc}<\overrightarrow{A}>$$since $\rho(\infty)=\rho(-\infty)=0$
$$E[\overrightarrow{\zeta}_\lambda]=-\frac{iq}{Mc}<\overrightarrow{A}>\neq0$$
That's the most interesting fact about our osmotic velocity, whose imaginary part possesses specific directionality. It exhibits highly unusual properties, leading us to suggest that it shares similarities with classical velocity rather than being solely a probabilistic diffusion term tied to the probability density (in chapter 5 we also observed its Newtonian hydrodynamical properties). As we have mentioned earlier, the imaginary component (as we've discussed earlier in chapter 3) enables us to define it as a variable that we refer to as "velocity due to diffusion." This interpretation highlights the departure from the conventional understanding of diffusion and emphasizes the unique nature of the osmotic velocity in our specific system.

It is worth exploring alternative perspectives and considering that the osmotic velocity could be the true velocity resulting from particle diffusion, going beyond a purely probabilistic interpretation. In this context, it becomes logical to propose that, during the interaction of our quantum fluid with the external quantum field, the particle may attain infinite velocities. Such a proposition aligns with Nelson's quantum theory and hidden variables, expanding our understanding of the phenomenon. It is important to note that every velocity field derived from the solution of Euler's or Schrödinger's equation represents a mean value, with numerous fluctuating velocity fields existing around them (\cite{bohm1989non}). In this scenario, $Im\left(-\overrightarrow{\xi}_B^0\right)$ represents the mean value, while $Im\left(-\overrightarrow{\xi}_B^\Lambda\right)$ represents the actual velocity resulting from diffusion. This perspective challenges conventional notions and encourages a deeper exploration of the complex dynamics involved.

A potential concern arises when considering bound states, where the gauge transformation $\overrightarrow{A}'=\overrightarrow{A}+\overrightarrow{\nabla}\Lambda$ could potentially lead to the "creation" of infinite probability densities. How can we explain that?

Initially, it is important to establish a clear definition of what the probability density truly represents. It is plausible to view it as an average density of the fluid or field under consideration. Consequently, regions with higher probability density indicate areas where a greater amount of fluid is present or passes through.
An objection may arise, suggesting that the probability density should reflect the likelihood of finding the point particle. In response, it is crucial to recognize that the particle is carried by the fluid itself. Therefore, it is logical to expect a higher concentration of the particle in regions where the fluid flow is more pronounced. In essence, the presence of greater fluid density naturally corresponds to a higher probability of locating the associated particle.

Furthermore, it is essential to recognize that the stochastic nature of quantum mechanics encompasses not only the velocities of the particle engaged in Brownian motion but also the fluctuations in the entire fluid that carries it. Throughout the presence of the fluid within the region bounded by the two cylinders, the actual fluid density undergoes continuous changes. Given our discussion regarding the actual velocities of the particle, it is reasonable to consider the existence of actual fluid densities "fluctuating" around the average fluid density, which corresponds to the probability density as previously mentioned. This viewpoint aligns with the inherent fluctuating behavior inherent in quantum mechanics, encompassing both particle and fluid dynamics. Due to the gauge we have that $v'=|m+\lambda'|$ where $\lambda'=\lambda\pm \delta\lambda$. So, we can say that $v'=v\pm \delta v$. The states that we obtain have the form
$$\rho_{fluid}(r)=\rho_{v\pm{\delta}v,n}(r)=N_{v\pm{\delta}v,n}^2\left(J_{v\pm{\delta}v}\left(\frac{\tau_{v\pm{\delta}v,n}}{d}(r-\alpha)\right)\right)^2$$
These are actually the fluid densities that appear during the motion of the fluid between the two cylinders. However, as we said before the average fluid density is the traditional probability density we extracted before. Specifically, we have that
$$<\rho_{fluid}(r)> = <\rho_{v\pm{\delta}v,n}(r)> = \rho_{v,n}(r)$$

It is important to note that our system is not unique in the presence of quantum fluctuations. Quantum fluctuations are inherent to various systems, although they may not be mathematically observable and are often referred to as hidden variables. This raises the question of why these fluctuations become apparent in our particular system. The answer, in essence, is straightforward. Classical electromagnetism provides us with the equation $$\overrightarrow{J}_{el.current}=\frac{1}{\mu_0}\overrightarrow{\nabla}\times \left(\overrightarrow{\nabla}\times \overrightarrow{A}\right)=\frac{1}{\mu_0} \overrightarrow{\nabla}\times \left(\overrightarrow{\nabla}\times \overrightarrow{A}'\right)$$ that helps us realize that electromagnetic theory is actually "aware" of Nelson's quantum mechanics and hidden variables. Let us remind you that what this actually means. It means that a given external field (in this occasion quantum field of the electric current) can cause many possible results (we can have many vector potentials according to the gauge). So, since classical electromagnetism "believes" in hidden variables and hidden quantum fluctuations, it's very logical that we will get them mathematically in quantum mechanics. 

In the case of wave packets we explained that the probability density is not exactly the fluid density.The same thing also holds for a bounded particle inside a force field. For instance,let's study the case where we have a scalar potential $V(x)=kx$ for $x>0$ and $\infty$ for $x<0$. In this case,  the solution to that equation according to \cite{wheeler2002classical} will be 
$$\rho_n^{(Airy)}(x)=\frac{\pi}{\sqrt{-z_n}}\left(Ai\left(\frac{(2mk)^{1/3}}{\hbar^{2/3}}x+z_n\right)\right)^2$$
where $z_n$ the $n$-th root of the Airy function.
Here, similarly to the Gaussian wave packet, the probability density is mass-dependent.
If we study the above equation, we find that the smaller the mass the more "diffused" the probability density seems to be.

We have mentioned that the field always behaves the same way regardless of our particle and its mass.But the trajectories of the particle inside this field will depend on its mass. So, for small masses in our example we have greater probability of finding the particle very far away from $x=0$. That's a region not so easy for classical particle to approach, due to the fact that it doesn't have enough energy to reach these regions.But everything is possible when the mass is small (the energies will be greater). We know that

$$E_n=-z_n\left(\frac{\hbar^2k^2}{2m}\right)^{1/3}\sim \frac{1}{m^{1/3}}$$

Quantum fluctuations can easily transfer the particle to those regions. We can say the same thing in the case of tunneling effect or in case of other bound states(it's easy to show for a particle in a potential $V(x)=\frac{1}{2}kx^2$).

Also, we must say that the more restrictive the potential is for the particle, the more intense the quantum fluctuations(or collisions of hypothetical particles). So, there is greater influence to the motion of the particle by those quantum fluctuations (and thus its average energy), and of course, this implies that there is greater dependence of the stochastic motion of the particle by its mass. That's why in a system where we have a scalar potential $V(x)=\frac{1}{2}kx^2$ for $x>0$ and $\infty$ for $x<0$ we have $E_n=(2n+\frac{3}{2})\hbar \sqrt{\frac{k}{m}}\sim \frac{1}{m^{1/2}}$ $(n=0,1,2...)$, while for a particle in a box of length $L$ (extremely restrictive system) we have $E_n=\frac{n^2\pi^2\hbar^2}{2mL^2} \sim \frac{1}{m}$ $(n=1,2,3...)$.

Returning to our previous discussions, it is important to emphasize that we cannot generally equate the probability density with the mean field density. However, in our specific system (or for a particle in a box), we can indeed make this identification. This is due to the fact that the probability density remains unaffected by the mass of the particle, as there are no force fields acting on the particle's motion between the two "walls." Consequently, the particle is effectively free to move within the entire region, leading to the association between the probability density and the mean field density in this particular scenario.

\section{Conclusions}

In conclusion, our focus has primarily been on bound states, attributing the Aharonov-Bohm effect to the rotational diffusion (the imaginary component of diffusion) rather than a change in states. Moreover, the osmotic velocity field serves as the fundamental quantity that gives rise to the probability distribution in the system. This observation holds true for both wave packets and bound states, as discussed in the final chapter. In these cases, the movement of the particle depends on its response to the quantum fluctuations of the field, with the resulting mass-dependent behavior reflected in the probability distributions. In essence, the osmotic velocity emerges as the most appropriate quantity for describing the stochastic motion of the particle under the influence of a fluctuating field. Consequently, working with diffusion becomes more significant than focusing solely on probability densities derived from it. Besides, the greater significance of osmotic velocity fields over probability densities was also supported by Nelson \cite{bohm1989non}. According to him, a probability distribution arises due to the existence of an osmotic velocity field.

This physical quantity, the osmotic velocity, arises from the interaction between our quantum field and another field, which can be mathematically derived from a scalar potential. This derivation is particularly relevant in the case of bound states, where it yields consistent physical results. The field in question is none other than the quantum field associated with the current. We argue that this interaction is only possible if the field is physical, supported by the fact that the body experiences Newtonian influence from the field in which it moves.

We observe that the rotational diffusion is more pronounced near the current, both inside and outside the cylinder where a magnetic field is present. This observation justifies the direct interaction between the two fields. We find common properties between quasi-current velocity, quasi-diffusion, and their traditional counterparts, while also we highlight the distinctiveness of rotational diffusion, which helps explain the differences between mechanical angular momentum and the quantized canonical angular momentum.

We further explain that the contribution of diffusion to the current velocity is also present in wave packets, such as Gaussian wave packets. In these cases, we attribute diffusion to the fluid-mechanical behavior of the field and the mass within it, reinforcing its physical nature. Additionally, we recognize that a physical system cannot be fully explained solely by its present state or the external field within it. The motion of a particle may depend on the past state of the fluid. For instance, in the case of diffusion in Gaussian wave packets, we attribute the phenomenon to quantum fluctuations of the field resulting from its interaction with another field, specifically a past state characterized by an harmonic oscillator field. This interaction led to the appearance of the diffusive term in the expression for the current velocity.

It is conceivable that the additional diffusion observed contributing to the current velocity may have originated from a past state in which the fluid was not confined by potential walls, and thus unable to interact with the electric current, but instead extended to the current region. We could also consider the past interaction of our field with the current field through our scalar potential, which extends from a to infinity. This interaction may have ceased to exist at some point and its effects become apparent in a new system, where we observe the abolishment of the force (in the case of wave packets) or the placement of potential walls(in the case of our system). However, it is important to note that comparing these systems may not be entirely accurate since one involves the time evolution of a state while the other involves time-independent states.

Furthermore, we emphasize that the fundamental field driving these phenomena is the quantum field under discussion, rather than the magnetic field. We can derive the force as a fluid-mechanical force acting on a body within a rotational vortex. In our framework, we consider the magnetic field as a turbulence constant that assists us in calculating the vorticity field. The reason why we observe a dependence of body motion on the magnetic field, even outside its direct influence, is because the magnetic field encompasses the electric current, which serves as the carrier of the quantum field. 

Next, we briefly touch upon time-dependent states (without conducting an exhaustive analysis) and aim to establish a connection between diffusion (specifically its imaginary part) and the phase of the wavefunction within a region influenced where vector potential seems to exist.

In the final chapter, we emphasize that the same principles that apply to the magnetic field also hold true for the vector potential. The vector potential serves as a tool that facilitates the calculation of the velocity due to diffusion, particularly evident in its expression as a function of vorticity for a fluid within and outside the magnetic field area. Rather than asserting that various vector potentials, while keeping the magnetic field constant, generate different physical outcomes in quantum physics, implying their fundamental nature, we take a different approach, drawing inspiration from Nelson's quantum theory and exploring the possibility of infinite osmotic velocities. We capitalize on the peculiar properties of the osmotic velocity to reconsider its physical interpretation in our particular system.

We propose that the imaginary component of the osmotic velocity corresponds to the actual velocity resulting from diffusion in bound states. This actual velocity continuously fluctuates around an average velocity(caused by diffusion) due to the inherent stochastic nature of the process. In time-dependent states, each experiment leads to distinct time-dependent diffusion, thereby influencing the phase of the wavefunction.

By delving into the unconventional properties of the osmotic velocity, we seek to offer a fresh perspective on the physical implications of this quantity. Our exploration opens up avenues to explore the fluctuating nature of diffusion in bound states and its impact on the wavefunction's phase in time-dependent scenarios.

To further clarify, in the context of bound states, we recognize that the velocity due to diffusion can be attributed to a specific moment in time. Within the bound state system, the multitude of infinite diffusion velocity fields corresponds to fluctuations around a particular value. These fluctuations give rise to a range of possible diffusion velocities, all coexisting within the system.

Expanding upon this notion, we extend our reasoning to address the issue of infinite probability densities associated with specific values of $m$ and $\lambda$. We posit that the probability density resulting from the gauge transformation of the vector potential is, in fact, a representation of the actual density of the fluid or field in which the body is immersed. This perspective allows us to understand the probability density as an expression of the tangible physical density within the system, accounting for the intricate dynamics of the quantum field.

By considering the temporal specificity of diffusion velocities and the physical interpretation of probability densities, we can also strive to elucidate the complex nature of quantum phenomena and their connection to the underlying fluid or field dynamics.

However, it is possible to introduce a scalar potential within the region between the two cylinders in such a way that the derived probability densities become dependent on the mass of the system. While this may initially appear as a challenge, in fact, it does not cause us any problem.

In this scenario, the probability densities can again be understood as snapshots, representing a particular moment in time. However, in this case, we must conceive that a probability density at a given instant does not solely arise from the density of the physical fluid at that moment, but also takes into account the body's response to the fluid motion, which is influenced by its mass. Hence, the derived probability densities are influenced by both the fluid density and the body's ability to follow the fluid's motion, modulated by its mass.

Nevertheless, we can confidently assert that for every moment in time and its corresponding fluid density, there exists a corresponding probability density. Thus, the presence of a fluctuating fluid gives rise to the generation of "fluctuating probability densities" around the conventional or average probability density. This fluctuation captures the inherent variability and dynamic nature of the quantum system, encapsulating the interplay between the fluid motion, mass, and resulting probability densities.

Alternatively, if we prefer a simpler explanation without delving into the intricacies of fluid densities, we can focus on the generation of osmotic velocity fields during quantum fluctuations. These osmotic velocities, in turn, give rise to infinite probability densities, as we know that every probability density arises due to the existence of an osmotic velocity field (we shall not forget about the Gaussian wave packet, where diffusion is actually responsible for the time evolution of our probability density, or the bound states, studied in chapter 8, where we established the role of quantum fluctuations and the diffusion of the body, caused by them, in the formation of probability distributions).

It is important to note that this perspective reveals a new insight. We observe that not only does the probability density originate from diffusion, but in the case of time-dependent states, the wavefunction itself is affected, acquiring an additional phase. Remarkably, while the probability density remains unchanged, this new phase emerges, highlighting the multifaceted nature of quantum phenomena and the interplay between diffusion, osmotic velocities, and wavefunction dynamics.

Furthermore, it is important to acknowledge that the infinite probability densities we derive as solutions do not necessarily have to be in the form of Bessel functions. The appearance of Bessel functions occurs only when the parameter $\Lambda$ takes the form $\sigma\theta$, where $\sigma$ is a constant. However, for other values of $\Lambda$, we can obtain a wide range of different functions as solutions, provided they are square-integrable. It is crucial to note that all these probability densities fluctuate around a particular probability density, highlighting the variability and diversity present in the system.

It is worth noting that previous discussions have arisen regarding the multi-valuedness of wave functions, particularly in the context of the Aharonov-Bohm paper. These discussions have explored the stochastic interpretation of quantum mechanics (\cite{wallstrom1989derivation},\cite{davidson2020multi}). However, we present a different approach to this topic, placing emphasis on the greater significance of diffusion comparing to the probability densities and its peculiar properties. By considering these properties, we can make logical assumptions that aid in explaining the phenomenon at hand. This alternative perspective offers a fresh viewpoint and may contribute to the ongoing discourse surrounding the interpretation of quantum mechanics.

We propose that electromagnetism may possess an underlying awareness of hidden variables, which could explain the manifestation of real diffusion and probability distributions. This idea aligns with the notion that classical electromagnetic phenomena fundamentally stem from the quantum realm. By recognizing the quantum nature of electromagnetic interactions, we can better understand how hidden variables may come into play, shedding light on the observed diffusion phenomena and probability distributions. This viewpoint also underscores the deep interconnection between classical and quantum physics in the realm of electromagnetism.

Our intention is not to provide a comprehensive and exhaustive explanation of this phenomenon. Rather, we aim to present some ideas and insights derived from Nelson's quantum theory and our hydrodynamical perspective. By exploring these perspectives, we hope to contribute to the understanding and interpretation of the observed phenomena, highlighting the significance of diffusion, probability densities, and the interplay between quantum and classical physics. It is through such explorations and discussions that new avenues for research and further investigation can be pursued.

In conclusion, our investigation of the Aharonov-Bohm phenomenon utilizing Nelson's quantum mechanics not only enhances our understanding of this specific phenomenon but also offers insights into the broader realm of quantum mechanics. A key concept that emerges from our analysis is the importance of past states in influencing present states. We have also highlighted the novel connection between the imaginary part of the osmotic velocity and the phase, which sheds new light on the dynamics of time-dependent states. While conventional systems, particularly bound states, do not exhibit a direct link between the phase and osmotic velocity, this connection becomes apparent in the case of time-dependent states, where the influence of the past is consistently accounted for. A prime example illustrating this concept is the behavior of a Gaussian wave packet. These findings not only contribute to our understanding of the Aharonov-Bohm effect but also deepen our comprehension of quantum mechanics as a whole.

The initial wavefunction has the following form 
$$\Psi_g(x,0)=\left(\frac{2}{\pi}\right)^{\frac{1}{4}}\frac{1}{\sqrt{\pi}}e^{-x^2/\alpha^2}e^{ik_0x}$$
while the time-evolved state according to Mita is the following
$$\Psi_g(x,t)=\left(\frac{2}{\pi}\right)^{\frac{1}{4}}\frac{1}{\sqrt{\pi}}e^{-(x-u_{0}t)^2/\epsilon^2}e^{i(k_0x-\frac{\hbar k_0^2}{2m}t+\delta)}$$
where $\delta=\frac{(x-u_{0}t)^2}{\epsilon^2}\frac{t}{T}-\frac{1}{2}tan^{-1}\left(\frac{t}{T}\right)$

It's easy to observe that the additional phase factor $\delta$ is related to the dispersive velocity via the relation
$$\overrightarrow{\xi}_g(x,t)=\frac{\hbar T}{mt}\frac{d\delta}{dx}\overrightarrow{e}_x$$

It is evident that the diffusion phenomenon diminishes significantly for large time intervals ($t \gg t_{\text{max}}$). This behavior can be readily understood within the framework of diffusion. As we previously discussed, diffusion characterizes the propensity of particles in a fluid (for instance) to move towards regions of lower concentration. This process relies on the existence of a concentration gradient between specific spatial points. However, as infinite time elapses and the concentration difference diminishes, with previously unoccupied regions becoming occupied, there is no longer a substantial reason to describe the fluid particles as exhibiting high diffusivity. Consequently, the diffusion process becomes less pronounced and eventually negligible for sufficiently large time intervals.

The quantum field we have been discussing exhibits dual properties of both wave and fluid mechanics. This dual nature can be observed in the equation for the current velocity, as presented in Chapter 3, specifically in the context of Gaussian wave packets. In that equation, we observe the presence of two terms: the velocity arising from wave propagation and the diffusion term, which indicates the fluid-mechanical behavior of the field.

Furthermore, these two characteristics are also evident in the phase of the wave function. The resulting phase is determined not only by the field's existence in the present state or field configuration but also by its existence in past states. This implies that the dynamics of the quantum field and its associated phase are influenced by the history of the field, incorporating information from previous states.

Therefore, the quantum field possesses a unique temporal aspect, where the past states and interactions shape the current state of the field. This temporal influence is crucial for understanding the nature of the field and its interplay with wave phenomena and fluid-mechanical behavior.

Indeed, the phase of the wave function retains information about the wave motion, which is related to the propagation of the wave. This term was present at t=0 when we eliminated the harmonic oscillator field. However, since the wave motion of the field, and consequently the particle's response to it, does not depend on potential interactions with the harmonic oscillator field, we can propose that the wave motion potentially existed throughout the entire time that the particle and its field were in the harmonic oscillator field.

In a broader sense, the phase of the wave function can be seen as a reflection of the past states of our field. It carries information about the history and interactions of the field, even beyond the current state or field configuration. This perspective aligns with our earlier discussions regarding the "creation" of an inhomogeneous quantum force from an external force field, which suggests that past interactions have a significant impact on the present state.

Applying this understanding to the Aharonov-Bohm effect, we can infer that the additional phase factor obtained in this scenario also has its origins in past interactions. This notion supports the idea that the interaction between the field and the current occurred before the confinement of the field in a region devoid of magnetic field. It provides a plausible explanation for the field's interaction with another field, despite the absence of a direct force-carrying interaction within the confined region.

By considering the temporal aspect and the influence of past states and interactions, we gain valuable insights into the nature of the field and its interactions, ultimately deepening our understanding of phenomena like the Aharonov-Bohm effect.

To strengthen our argument regarding the relationship between additional phase factors, diffusion, and past field interactions, it is pertinent to refer to a well-known expression for a free particle, which is $\Psi_{\text{free}}(x,t) = A e^{i(k_0 x - \frac{\hbar k_0^2}{2m} t)}$. It is important to note that this wave function is non-square integrable, rendering it unsuitable as a valid solution to the time-dependent Schrödinger equation. However, even in this case, we can observe that the Gaussian wave packet exhibits an additional term, denoted as $\delta$, which is closely related to the osmotic velocity.

Despite both systems being associated with free particles, the presence of the past field interactions becomes apparent in one of them. This observation highlights the significance of considering the influence of past interactions in understanding the behavior of quantum systems. By recognizing the role of these interactions, we can unravel the connection between additional phase factors, diffusion phenomena and the historical evolution of the quantum field.

We can also observe the parallelization of these two systems in another way by writing the equation for the current velocity for the Gaussiam wave packet as follows 
$$\overrightarrow{\eta}_g(x,t)=\overrightarrow{\eta}_{free}(x,t) +\frac{mt}{T}\overrightarrow{\xi}_g(x,t)$$

The above equation is analogous to eq. (\ref{eq:19}). On the left-hand side, we have the current velocity associated with a free particle when there has been a previous field interaction. On the right-hand side, we have two terms. The first term represents the current velocity for a free particle where there has been no previous field interaction (we know that $\rho_{free}(x,0)$ represents the solution for the time-independent Schrödinger equation in a free-field area, while also that $\overrightarrow{\xi}_{free}(x,t)=0$). The second term contains the dispersive velocity, which arises due to the interaction between the fluid and the harmonic oscillator field. This additional term captures the influence of the past field interactions on the particle's behavior and contributes to the overall motion of the system.

It is important to note that in this specific example we are using for parallelism, there is also a difference in the modulus of the wave function between these two systems. However, in the case of the time-dependent Aharonov-Bohm effect, there is no change in the probability density. Despite this difference, our main focus here is on the phase and the velocities associated with it.

This analysis also sheds light on the reason why, in a force-free region, solutions to the Schrödinger equation ($\Psi_{free}$) exhibit wave-like behavior (the well-known wave aspect has been demonstrated by experiments like the double-slit experiment), despite the Schrödinger equation being a diffusion equation. The field can be likened to water, which is both a fluid and capable of supporting wave propagation on its surface. More specifically, this might be the reason why the Schrödinger equation is a diffusion equation, while the solutions to it are complex-valued functions (due to this dual behavior of the quantum field). However, if the mass of a particle is large, it becomes less responsive to the oscillatory behavior of the "fluid particles" and the transfer of waves. As a result, wave-like or fluid-like behavior is not prominent in the motion of very large bodies, as evident from the Gaussian wave packet where the parameters $u_0$ and $\xi_g$ approach zero for large masses.

Our analysis of the influence of past interactions on our fluid states allows us not to adopt other unsuccessful hydrodynamic approaches to the phenomenon. One of these was mentioned by \cite{philippidis1982aharonov}, \cite{strocchi1974proof} and \cite{zuchelli1984inadequacy} and pointed out that possibly despite the fact that wavefunction is zero in the magnetic field region there is a small amount of this "mathematical" fluid that can enter this region. Of course this suggestion is largely recognized by the authors themselves as unrealistic. We actually share this thinking. As we have repeatedly said, a possible interaction does not depend on the range of the wavefunction, but on the state of the physical field in the past and the resulting imaginary part of the osmotic velocity.

To sum up, we must mention that, even in the case that our description regarding the effect is not compeletely consistent with physical reality, the leading role in change of states is taken by the imaginary component of the osmotic velocity-it is in fact the second term of the mechanical momentum divided by the mass-and not by the vector potential. This statement is in agreement with Nelson and our previous analysis about the importance of osmotic velocities in "creating" time dependent and bound states. Anyway, we insist that a quantum description is required for the exploration of this effect, that takes into consideration the quantum background field of the current and the charge.

{}


\begin{thebibliography}{}


\bibitem{ahaBohm1959} Y. Aharonov and D. Bohm, Significance of Electromagnetic Potentials in the Quantum Theory Physical Review {\bf 115} 485 (1959).

\bibitem{madelung} Quantum theory in a hydrodynamical form Zeitschrift für Physik {\bf 40} 322 (1927)

\bibitem{takabayasi1} Hydrodynamical formulation of quantum mechanics and Aharonov-Bohm effect Progress of Theoretical Physics {\bf 9} 1323 (1983)

\bibitem{mita2003dispersive} Dispersive properties of probability densities in quantum mechanics American Journal of Physics {\bf 71} 894--902 (2003)

\bibitem{nelson1966derivation} Derivation of the Schr{\"o}dinger equation from Newtonian mechanics Physical review {bf\ 150} 1079 (1966)

\bibitem{berry1979nonspreading} Nonspreading wave packets American Journal of Physics {\bf 47} 264--267 (1979)

\bibitem{mita2021schrodinger} Schr{\"o}dinger's equation as a diffusion equation American Journal of Physics {\bf 89} 500-510
(2021)

\bibitem{philippidis1982aharonov} Aharonov-Bohm effect and the quantum potential Nuovo Cimento B;(Italy) {\bf 71} (1982)

\bibitem{arbab2011analogy} The analogy between electromagnetism and hydrodynamics Physics Essays {\bf 24} 254 (2011)

\bibitem{green2012fluid} Fluid vortices {\bf 30} (2012)

\bibitem{heifetz2015toward} Toward a thermo-hydrodynamic like description of Schr{\"o}dinger equation via the Madelung formulation and Fisher information {\bf 45} 1514--1525 (2015)

\bibitem{bohm1989non} Non-locality and locality in the stochastic interpretation of quantum mechanics Physics Reports {\bf 172} 93--122 (1989)

\bibitem{bohm1952suggested} A suggested interpretation of the quantum theory in terms of hidden variables. I Physical review {\bf 85} 166 (1952)

\bibitem{wallstrom1989derivation} On the derivation of the Schr{\"o}dinger equation from stochastic mechanics Foundations of Physics Letters {\bf 2} 113--126 (1989)

\bibitem{davidson2020multi} Multi-valued vortex solutions to the Schr{\"o}dinger equation and radiation Annals of Physics {\bf 418} 168196 (2020) 

\bibitem{wheeler2002classical} Classical/Quantum Dynamics in a Uniform Gravitational Field: An Unobstructed Free Fall Preprint, Reed College Physics Department August (2002)

\bibitem{becker2019asymmetry} Asymmetry and non-dispersivity in the Aharonov-Bohm effect Nature communications {\bf 10} 1700 (2019)

\bibitem{becker2017observation} Observation of quantum forces in the Aharonov-Bohm effect preprint (2017)

\bibitem{vaidman2012role} Role of potentials in the Aharonov-Bohm effect Physical Review A {\bf 86} 040101 (2012)

\bibitem{vaidman2015reply} Reply to “Comment on ‘Role of potentials in the Aharonov-Bohm effect’” Physical Review A {\bf 92} 026102 (2015) 

\bibitem{strocchi1974proof} Proof of the charge superselection rule in local relativistic quantum field theory Journal of Mathematical Physics {\bf 15} 2198--2224 (1974) 

\bibitem{zuchelli1984inadequacy} Inadequacy of hydrodynamical theories of the Aharonov-Bohm effect International journal of theoretical physics {\bf 23} 407--415 (1984)

\end{thebibliography}
\end{document}